# Generalized Neural Networks for Real-Time Earthquake Early Warning

Xiong Zhang[1*], Miao Zhang[2]

[1]Engineering Research Center for Seismic Disaster Prevention and Engineering Geological Disaster Detection of Jiangxi Province, East China University of Technology, Nanchang, Jiangxi 330013, China

[2]Department of Earth and Environmental Sciences, Dalhousie University, Halifax, Nova Scotia B3H 4J1, Canada

Corresponding author: Xiong Zhang (zxiong@mail.ustc.edu.cn)

**Abstract:**

Deep learning enhances earthquake monitoring capabilities by mining seismic waveforms directly. However, current neural networks, trained within specific areas, face challenges in generalizing to diverse regions. Here, we employ a data recombination method to create generalized earthquakes occurring at any location with arbitrary station distributions for neural network training. The trained models can then be applied to various regions with different monitoring setups for earthquake detection and parameter evaluation from continuous seismic waveform streams. This allows real-time Earthquake Early Warning (EEW) to be initiated at the very early stages of an occurring earthquake. When applied to substantial earthquake sequences across Japan and California (US), our models reliably report earthquake locations and magnitudes within 4 seconds after the first triggered station, with mean errors of 2.6-6.3 km and 0.05-0.17, respectively. These generalized neural networks facilitate global applications of real-time EEW, eliminating complex empirical configurations typically required by traditional methods.

Earthquake monitoring is a primary task in seismology, and reporting earthquake parameters in real time has long been a critical effort for earthquake early warning (EEW)[1-4]. The current EEW systems typically require a few seconds to 1 min after an earthquake occurs to issue warning information to public[3]. The systems commonly consist of several modules, including data processing, source parameter evaluation, and alert filtering, used to report earthquake alarms based on continuous waveform streams[5-7]. The alarm is triggered if specific conditions are met, such as distinguishing teleseismic events, triggering a certain number of stations, reaching a certain magnitude, and meeting a percentage threshold of triggered stations (e.g., 40%)[5,7]. Complex empirical threshold value settings are involved in each processing step, making it a challenge to define the optimal alert criteria in EEW systems. Implementing overly strict criteria, such as requiring too many or a large number of stations to trigger, can negatively impact the real-time efficiency of EEW systems, while loose criteria can result in false alarms[5]. Theoretically, earthquake parameter determination may require data from at least four triggered stations to ensure accuracy[5-7]. The magnitude determination often requires 3 s P arrivals for a single station[8-9]. Therefore, the time delay for issuing a warning is the duration from the origin time to 3 seconds after the last station triggers when multiple stations are used to estimate the magnitude. However, in real applications, the time delay for issuing a warning may be even longer due to malfunctions in stations or system delays[6,10]. Therefore, an efficient real-time monitoring algorithm should not only be computationally fast in a concise way without extensive empirical configurations but also be able to solve the earthquake parameters with limited data available from the triggered stations at the early stage of the earthquake occurring.

Traditional traveltime-based earthquake monitoring involves many steps, including earthquake detection, phase picking, phase association, earthquake location and magnitude evaluation[8-23]. With recent advancements in machine learning, the automatic construction of earthquake catalogs is now feasible without the need for manual intervention[21-23], which significantly enhances our ability to monitor and comprehend seismic activity. However, as the workflow involving many steps, the error in either of the steps would potentiality affect the final location and magnitude results, and they require more time to receive and analyze the earthquake signals. Unlike the automatic construction of earthquake catalogs, EEW systems require the rapid reporting of earthquake parameters at the very early stages of occurrence, often with limited information available from network stations. Thus, the location method based on the concept of "triggered and not-yet triggered stations" has been proposed to decrease the EEW "blind zone" which experiences the most significant damages[24-25]. This method utilizes both the arrival time picks from the triggered station and locations of the not-yet triggered stations to constrain the earthquake epicenter.

Another effort to monitor earthquakes without human intervention involves direct data mining from waveforms to detect earthquakes and evaluate parameters using deep learning techniques[26-37], bypassing many intermediate steps which require complex empirical settings such as various threshold values and alert criteria. This approach utilizes waveform features more than phase picks. For single-station monitoring, the epicentral distance and magnitude can be determined by mining waveform data with various networks[26-28], or earthquake locations can be roughly classified into several blocks within a specified region[29]. However, to obtain more accurate earthquake parameters, a network of stations may be necessary to provide sufficient constraints for neural network learning[30-37]. Zhang el al. (2020) demonstrate that the fully convolutional neural network is efficient to map the waveforms of fixed network stations to the earthquake location labeled by 3D Gaussian distribution function. van den Ende et al. (2020)

utilize a graph neural network to estimate the earthquake location and magnitude by incorporating the spatial information of the stations into the neural network input. The waveform features from each station are extracted independently and then combined by a component of multi-layer conception; thus the output is not sensitive to the station order. Nevertheless, the generalization of monitoring neural networks remains a challenge due to the diverse geometric distribution of stations or geological structures. These methods commonly require transfer learning when applied to new regions[30-37]. Furthermore, earthquake monitoring for early warning requires neural networks to handle both triggered and not-yet-triggered stations from the outset of an earthquake event[24-25], adding complexity to neural network learning. Zhang et al. (2021) extend the fully convolutional neural network to early warning applications, showcasing its capability to determine source parameters at the onset of an earthquake. However, the neural network's application is restricted to the same region and station distribution as the training set.

In this study, we develop data augmentation methods to train generalized neural networks capable of real-time earthquake monitoring and early warning in diverse regions with different station distributions. Despite the abundance of seismic datasets shared online within the seismology community[38], it remains a challenge to collect a sufficient number of training earthquakes with diverse station distributions and monitoring settings. In our application, we utilize a recombination method of station seismograms to create numerous generalized training earthquakes that occurred at any location within the study area with arbitrary station distributions. The trained neural networks are applied across diverse regions for real-time earthquake early warning, extracting features from limited available data at the early stages of earthquake occurrence, without complex empirical settings as required in traditional methods.

**Network models and data recombination**

We have designed three neural networks for earthquake detection, location, and magnitude estimation, with seismic data streams continuously fed into the network models (Fig. 1). The earthquakes are simultaneously detected and located with waveforms from multiple stations, and then the neural network estimates the earthquake magnitude for each station when an earthquake is detected. Our models are built upon fully convolutional networks with detailed configurations described in the Methods section. The outputs are labeled using 1D or 3D Gaussian distributions to represent the detected P arrival, location, and magnitude. We incorporate station XY coordinates as two channels of the network input and sort the input data by station X and Y coordinates, ensuring that a specified earthquake corresponds to a unique input with well-defined station order. To ensure a generalized neural network, the key technique involves the production of training samples by recombining station seismograms to create generalized earthquakes occurring within the study area with arbitrary station distributions (Fig. 2). Similar to the 1D layered velocity model approximation used in common location methods[17,39], we adopt the assumption that seismograms with the same epicentral distance and depth across various global areas exhibit similarity in phase arrival times. This assumption enables us to recombine collected seismograms and generate training earthquakes in arbitrary monitoring areas. The base training dataset comprises 94,586 single-station seismograms from real earthquakes that occurred in Italy, Oklahoma (US), and Southern California (US) (Fig. 3). We group the 94,586 station seismograms by the corresponding epicentral distance and earthquake depth (Fig. 2 and 3), and randomly recombine the seismograms to create 355,001 training earthquakes. The 3D Gaussian distribution outputted by the neural network represents an earthquake location area of 50 km × 100 km horizontally, while the station distribution covers the earthquake location range within a broader region of 82 km × 110 km. Therefore, in a typical monitoring setting, we assume that the generalized stations are distributed within a range of 0-82 km in the X direction and 0-100 km in the Y direction, while the generalized earthquakes occur within the monitoring area of 16-66 km in the X direction, 0-100 km in the Y direction, and 0-20 km in the Z direction. To generate a training sample, we randomly select an earthquake location ($s_x$, $s_y$, $s_z$) within the earthquake range and 4-12 station locations within the station range. We then calculate the epicentral distance $r$ and randomly select a three-component waveform with epicentral distance of $r$ and depth of $s_z$ from the grouped base dataset for a synthetic station. Fig. 2A illustrates a representative generalized training earthquake monitored by four stations with waveforms from two different real earthquakes in the base dataset. The E and N components of the waveforms are rotated to the new azimuth direction centered at the generalized earthquake location, allowing us to use phase azimuth to constrain earthquake location. To accommodate amplitude variations among different earthquakes, we filter the three-component seismograms within a frequency range of 1-9 Hz and normalize their amplitudes to an equivalent magnitude of $M_L$ 3.0. We achieve this normalization using Hutton and Boore (1987)'s empirical equation[40]: $M_L = \log(A) + 1.110\log(r/100) + 0.00189(r - 100) + 3.0$ . We also generated 2000 training earthquakes outside the monitoring areas to enhance the network's ability to distinguish abnormal events, which are labeled as zeros. To simulate the real-time EEW process, we shift the waveforms of the generalized earthquakes and vary the length of effective signals in the current time window (30 s) to generate training samples when only a subset of stations is triggered (see Methods section). The generalized earthquake data contain the crucial features related to earthquake location, although we omit many other physical factors, such as the focal mechanism and complex geological structures across various regions. The neural networks, trained with the generalized earthquakes, are subsequently applied to various regions (Fig. 3).

**Pseudo real-time monitoring of the main shocks in Osaka, Japan and Ridgecrest, US**

We apply the trained neural networks to the main shock ($M_{JMA}$ 6.1) that occurred on 18 June 2018 in Osaka, Japan and the main shock ($M_w$ 6.4) that occurred on 4 July 2019 in Ridgecrest, US (Fig. 4). The two earthquakes were monitored by 12 stations with different spatial distributions, however, many of which recorded clipped waveform - where the amplitude of the seismic signal exceeds the dynamic range of the recording instrument - potentially resulting in an underestimated magnitude, as shown in Fig. S1. Following our magnitude estimation process, we recalculated their magnitudes as $M_L$ 5.5 and $M_L$ 5.8, respectively, to facilitate consistent comparison with the predicted results of the neural networks. To simulate real-time monitoring, the continuous waveforms from the 12 stations are fed into the neural network with a 30 s truncating window and 0.5 s time interval along with the station locations. The waveforms are filtered from 2.0 to 8.0 Hz and sorted by the station coordinate increments in X and Y directions. The detection and location networks output the 1D and 3D probability density functions (PDFs) with Gaussian distributions simultaneously. A 1D detection label with a maximum PDF greater than 0.7 indicates an earthquake detected in the input time window, while a 3D location label with a maximum PDF greater than 0.6 indicates a well-located earthquake. The first triggered P arrival is identified by the maximum value in the detection label. The origin time is calculated by subtracting the travel time from the first triggered P arrival time. In addition, we calculate the epicentral distances and theoretical P arrival times to determine the triggered stations and estimate their magnitudes. If the theoretical P arrival time falls within the current time window, it indicates that the corresponding station is triggered. The magnitude neural network takes the epicentral distances and waveforms from the triggered stations as input, and outputs 1D Gaussian distributions that represent the normalized magnitudes which exclude the contribution of maximum amplitudes. The $M_L$ magnitudes are calculated as the addition of the normalized magnitudes and the logarithm of the maximum amplitudes. The predicted magnitudes are considered robust if the maximum PDFs exceed 0.6 and the earthquake signals last for more than 2 seconds (estimated by the difference between the current time and the theoretical P arrival time). The input waveforms are filtered from 1.0 to 9.0 HZ, and the final $M_L$ magnitude of the detected earthquake is the mean of the triggered stations.

The neural networks detect the earthquake signals and predict the earthquake parameters instantly with continuous waveform input (Movie S1 & S2). In Fig. 4, we present monitoring snapshots for the main shocks in Osaka, Japan, and Ridgecrest, US, captured at the 4th and 15th seconds after the first triggered station. At the 4th second, the epicenter errors are 3.9 km and 3.2 km with maximum location PDFs of 0.698 and 0.665 for the Osaka and Ridgecrest earthquakes, respectively. The final $M_L$ magnitudes are 5.3 and 5.8, calculated from the means of 3 and 5 triggered stations meeting the criteria of maximum PDFs exceeding threshold values and earthquake signals lasting over 2 seconds. Both the epicenter and magnitude errors remain similar to the results obtained at the 15th second, even though the latter results are evaluated with more data. The results imply that the monitoring system can reliably report earthquake parameters at the 4th second after the earliest P arrival. In fact, the main shocks could be detected and located as early as 2.2 seconds and 2.7 seconds, respectively, although the magnitudes are underestimated (M 4.8 and M 5.2), as shown in the supplementary materials (Fig. S2). This indicates that the first alarms could be reported even earlier.

**Robustness analysis for the generalized neural network models**

We apply the neural networks to 528 relatively large earthquakes and the continuous records of one day's duration in both Osaka, Japan, and Ridgecrest, US to test the robustness of the network models (see Fig. 5). The continuous waveforms are fed into the neural network, and the detection and location threshold values are set to 0.7 and 0.6. If both PDF values are satisfied, the predicted results enter the shortlist of an earthquake, and the input windows with similar origin times and similar locations are considered as the same event. We remove the duplicated events by choosing the result with the best location PDF and the effective signals occur within the range of 15~20 s, considering a reasonable length of waveforms for better parameter accuracy. We detect and locate 126 events on June 18, 2018, in Osaka, Japan, and 389 events on July 4, 2019, in Ridgecrest, US, distributed around the main shocks (Fig. 5). Although the neural networks are trained with M ≥ 2.5 earthquakes, they can also detect and locate earthquakes smaller than M 2.5 (Fig. 5E and 5F). To assess the accuracy of earthquake parameters predicted by the generalized neural networks, we utilize the waveforms of 179 events ($M_{JMA}$>1.0) occurred from 16 Jan to 25 Dec 2018, in Osaka, Japan and 349 events (M>3.5) occurred from 4 July to 16 Nov 2019, in Ridgecrest, US as inputs. The monitoring stations are the same as in the continuous data tests of the main shocks; however, some of the stations may not work or the data are missing for some events in one year range. We recalculated the events to $M_L$ magnitude using Hutton and Boore (1987)'s empirical equation[40] to compare with the neural network results. The epicenter errors for the Osaka and Ridgecrest earthquakes are 2.6 km and 4.5 km (Fig. 5A and 5B), respectively, and the magnitude errors are 0.08 and 0.12, respectively (Fig. 5C and 5D). The results show that the earthquake parameters can be evaluated without any human intervention and the trained neural network can be applied generally in different areas and monitoring projects.

**Performance of the neural networks for early warning**

To comprehensively test the performance of the neural networks, we utilize the 179 events in Osaka, Japan and 349 events in Ridgecrest, US to simulate the early warning progress and assess the errors of the predicted parameters at different times. We truncate the continuous waveforms of the events to different time windows with various lengths of effective signals and utilize the neural networks to predict the corresponding earthquake parameters (Fig. S3). Fig. 6 and 7 show comparisons between the predicted and cataloged earthquake parameters at the 4th and 15th seconds after the first triggered stations. At the 4th second, the epicentral distance and magnitude errors are 4.8 km and 0.17 for the Osaka earthquakes, and 6.3 km and 0.17 for the Ridgecrest earthquakes, respectively. By the 15th second, the epicentral distance and magnitude errors improve to 2.7 km and 0.04 for the Osaka earthquakes, and 4.9 km and 0.08 for the Ridgecrest earthquakes, respectively. More error statistics at different times show that the earthquake early warning can be activated as early as 3~4 seconds after the first triggered station, and the accuracy of earthquake parameters generally improves when the monitoring stations received more data (Fig. S3). However, depth resolutions are relatively lower with the monitoring stations at surface. The depth errors are 2.7 km and 5.0 km for Osaka earthquakes, and 4.1 km and 3.1 km for Ridgecrest earthquakes at 4th and 15th seconds, respectively. When the effective signals are increased to 25 seconds, the depth errors in Japan earthquakes are perturbed, as shown in Figure S3. This implies that the latter complex scattered waves may be challenging to learn due to various geological structures in different areas.

**Generalization ability of the neural networks**

We test the generalization ability of the neural networks by applying them to different regions across Japan and North California, US (see Fig. 8). One-hour continuous data (starting 90 seconds before the earthquake occurs) of 130 relatively large earthquakes, monitored by varying numbers of stations, are input into the neural networks. This includes 57 earthquakes from Japan and 73 earthquakes from North California, US, respectively. We set the monitoring areas for the 130 earthquakes according to the downloaded station distributions, and ensure that the monitoring areas cover all the stations. We position the center of the monitoring area around the mean values of station coordinates to include as many stations as possible. If the number of stations exceeds 12, we use the closest 12 stations to the epicenter to construct the monitoring area. In the case of Japan earthquakes, most are monitored by 11-12 stations, as shown in Fig. 8, while in Northern California, US, most earthquakes are monitored by 5-8 stations. Fig. 8 C, D and E show the error statistics of the predicted epicentral distance, depth and magnitude for the 130 relatively large earthquakes. The mean errors of the epicentral distance, depth, and magnitude are 4.9 km, 4.0 km, and 0.1 for the 130 relatively large earthquakes, respectively (Fig. 8). Most of the epicentral distance errors fall in the range of 2-6 km, and the magnitude errors are less than 0.5. However, the depth errors are relatively large in the range of 0-5 km possibly due to the weaker constraints imposed by surface stations, similar to traditional methods. We detect 4,385 earthquakes from all of the one-hour continuous datasets. The results reveal that a large number of small earthquakes are detected, even though our neural networks are trained with relatively large earthquakes (M≥2.5) (Fig. 8F).

**The first alarms in real-time monitoring with different station distributions**

We analyzed the monitoring results for 130 relatively large earthquakes, focusing on the first alarms where both detection and location PDF values satisfied threshold criteria of 0.7 and

0.6. Fig. 9 illustrates that most first alarms could be issued with approximately 4 seconds after the first triggered P arrivals. The first alarm time ($T_f$) after the first triggered P arrivals varies based on station number and distributions. For earthquakes monitored by dense station distributions, $T_f$ is relatively small (~4 seconds), while for those monitored by sparse station distributions (3-5 stations), $T_f$ may exceed 8 seconds. Despite the varying $T_f$, our results indicate that at the first alarms, the accuracy of earthquake parameters is sufficient for early warning (Fig. 9). The mean errors for epicentral distance, depth, and magnitude are 4.9 km, 4.0 km, and 0.2, respectively. The errors in estimated earthquake parameters are similar across cases with different station numbers. Neural networks may require longer effective signals for cases with sparse station distributions. In contrast, dense distributions offer more triggered stations, allowing for better parameter constraint at the earthquake's onset, resulting in earlier first alarms. The results suggest that neural networks can effectively extract features from various number of stations and report alarms instantly when current available waveforms meet the minimum requirement for constraining the earthquake parameters.

**Discussion**

**Potential for global real-time EEW application**

We trained the neural networks using data from Italy, Oklahoma, and Southern California. They can now be applied for real-time earthquake early warning in different regions without additional training, offering broad applicability to reduce seismic hazards globally. The training data from Southern California were generated before occurrences of the testing earthquakes in Ridgecrest. Despite the temporal difference between the training and testing earthquakes, the geological structures and waveforms may exhibit similarity, given that Ridgecrest is geologically part of Southern California. Nevertheless, the monitoring errors for Ridgecrest earthquakes are similar to the results observed for earthquakes in Japan and Northern California which have no spatio-temporal overlap. This suggests that the neural networks have learned generalized features for earthquake detection and parameter evaluation, rather than specific features of a particular local area. These models offer a straightforward way to monitor earthquakes in any region without complex empirical settings. Although threshold values for Gaussian distribution PDFs need to be set, the same values can be used across regions without considering station distributions and monitoring settings. The generalization ability facilitates the deployment of earthquake early warning systems. Moreover, the neural networks can mine the real-time waveform streams to the utmost and report alarms automatically.

**Potential improvements**

We make use of the move-out features represented by the P arrival times from sorted stations, and utilize the convolutional neural networks to extract the earthquake location features from both waveforms and station coordinates. We avoid extracting features separately for each station, as is done in the graph neural network[33,37]. This decision is based on the consideration that the images represented by the sorted waveforms exhibit similarity for earthquakes with similar locations, implying a strong constraint on earthquake locations. The convolutional layer is indeed a powerful tool for extracting move-out features between stations. However, the neural networks could potentially benefit from various structures. Integrating convolutional layers for extracting local features and transformer structures for capturing overall features seems like a promising approach[41]. This combination could potentially enhance the neural network's ability to learn both specific details and global patterns, providing a more comprehensive understanding of earthquake features.

The monitoring system may not be able to distinguish multiple events occurred in the same input window of 30 s (Fig. S4). The neural networks may output the results of the largest earthquake with dominating signals in the input time window, potentially missing the small earthquakes. In another aspect, if multiple events with similar magnitude are mixed in the same window, the outputted PDFs may be very small and the earthquakes could not be distinguished, or the output earthquake parameters from different events are mixed, which may result in false alarms of the earthquakes. This poses a challenge in earthquake monitoring for small earthquakes, even though the large earthquakes rarely occur at the same time in a monitoring area. A potential solution to this problem is to stack two earthquake samples together to simulate cases with two earthquakes occurring nearly.

Onsite early warning methods commonly use 3-4 second wavefoms from a single station to assess local ground motion or magnitude[42-43]. However, for a robust system, the confirmation of the alert would require several triggered stations, and the early warning would be activated until the last station is triggered. Consequently, the time elapsed could be greater than 3-4 seconds after the first station triggered. Additionally, the teleseismic signal poses a challenge when relying on a single station for EEW. Traditionally, teleseismic signals can be distinguished based on frequency features, as the high-frequency energy is significantly attenuated for teleseismic signals[7]. In our application, we generate 2000 earthquakes outside of the monitoring ranges as abnormal events to train the neural networks. Earthquakes outside of the monitoring areas exhibit very different features from normal events in terms of the waveforms from multiple stations. Incorporating more teleseismic waveforms for training could further enhance the efficiency of the network models. The neural networks may be further generalized by incorporating other information for training. For instance, smartphones, which cover high-population regions, have been proven to be applicable for EEW, and signals from accelerometers in smartphones could be utilized to detect earthquakes[44].

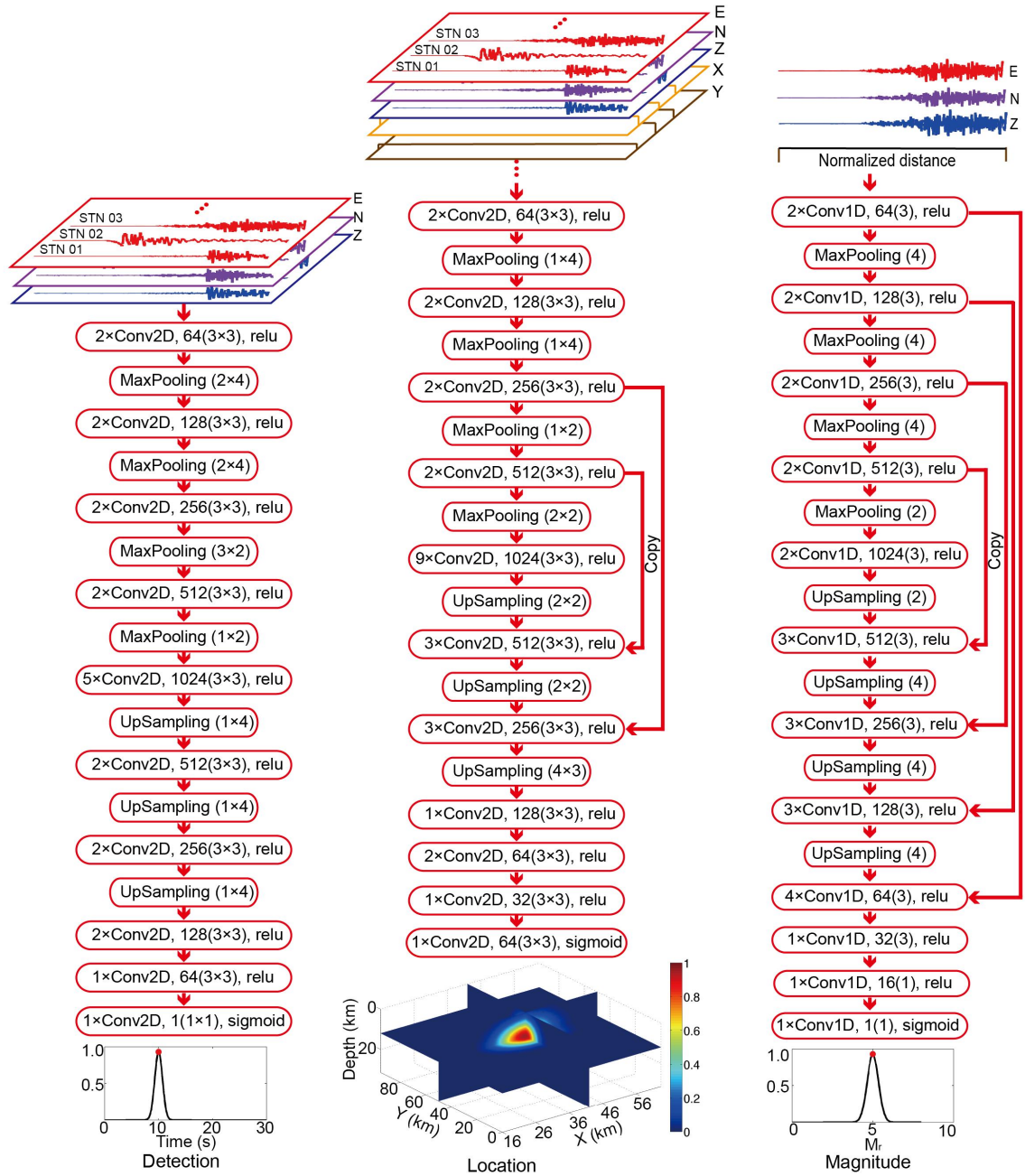

**Figure 1.** Inputs, outputs, and neural networks. The three component waveforms or station XY locations are fed into the neural networks, and the outputs are labeled as 3D or 1D Gaussian distributions to represent the earthquake parameters. The convolutional layers read as number of kernels (kernel size).

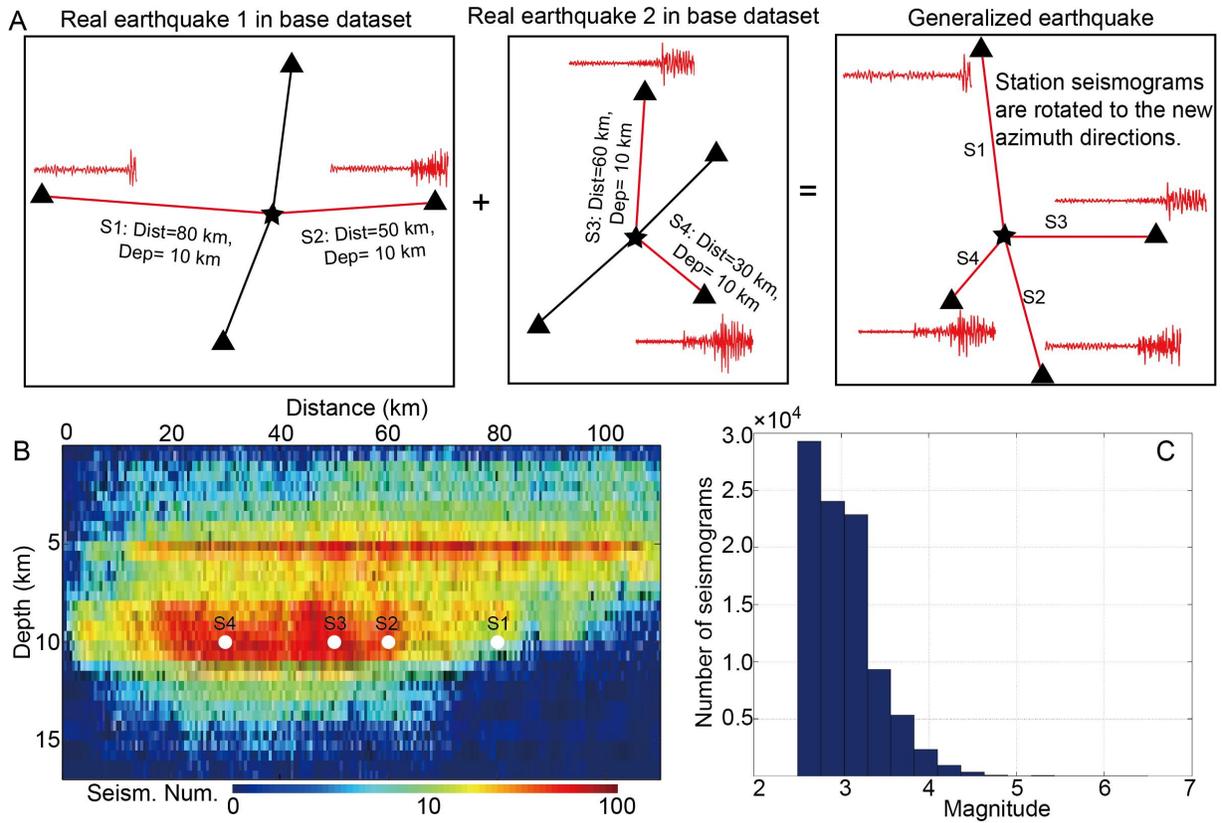

**Figure 2.** Data recombination with the base dataset. (A) A typical generalized earthquake produced by recombining seismograms from two real earthquakes; the earthquake locations are represented by stars, and monitoring stations with seismograms are shown as triangles. (B) The base dataset consisted of 94,586 seismograms grouped by epicentral distance and earthquake depth. Four single-station seismograms (S1, S2, S3, S4) in panel A are extracted from the base dataset and marked as white dots. (C) The magnitude statistics for the base dataset.

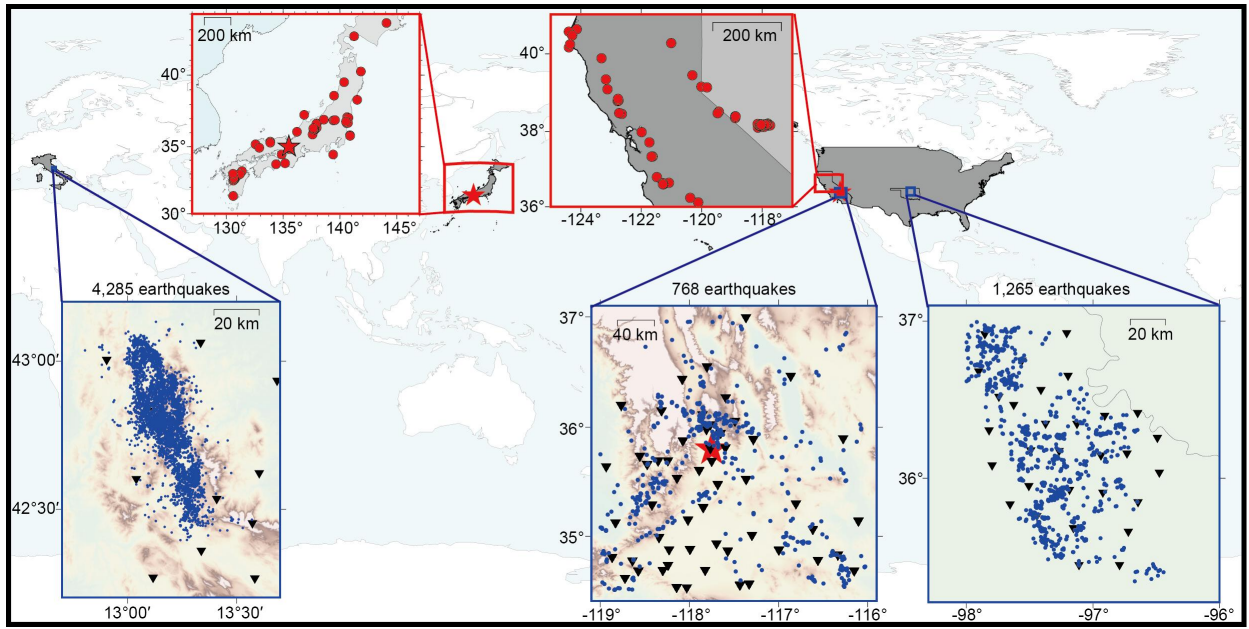

**Figure 3.** The training data and testing sites. The base training dataset includes 94,586 single-station seismograms from earthquakes in Central Italy, Southern California (US), and Oklahoma (US), indicated by blue dots; the triangles represent the stations. The red stars denote the testing sites for earthquake sequences in Osaka, Japan, and Ridgecrest, US, while the red rectangles indicate larger testing areas including 130 relatively large earthquakes (red dots) across Japan and Northern California (US).

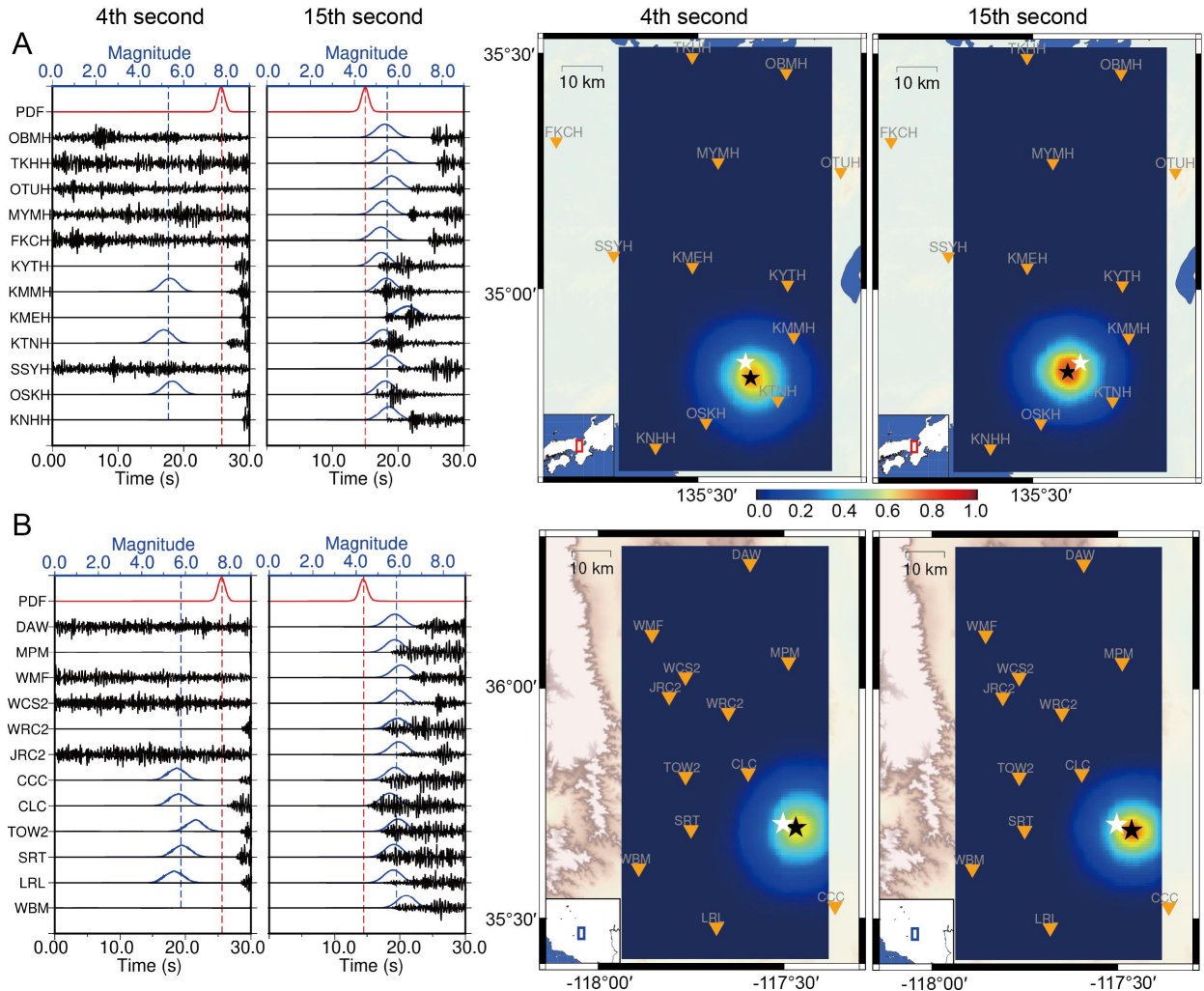

**Figure 4.** Real-time monitoring of two main shocks. (A) The earthquake ($M_{JMA}$ 6.1 or $M_L$ 5.5) occurred on June 18, 2018, in Osaka, Japan. (B) The earthquake ($M_w$ 6.4 or $M_L$ 5.8) occurred on July 4, 2019, in Ridgecrest, US. The monitoring snapshots at the 4th and 15th seconds after the P arrival of the first triggered stations are shown in the figures. The black, blue, and red curves represent the input waveforms (Z components), magnitude PDFs, and detection PDFs, respectively; the blue dashed lines mark the mean of the predicted magnitudes; the black and white stars in the right figures are the predicted and cataloged earthquake locations.

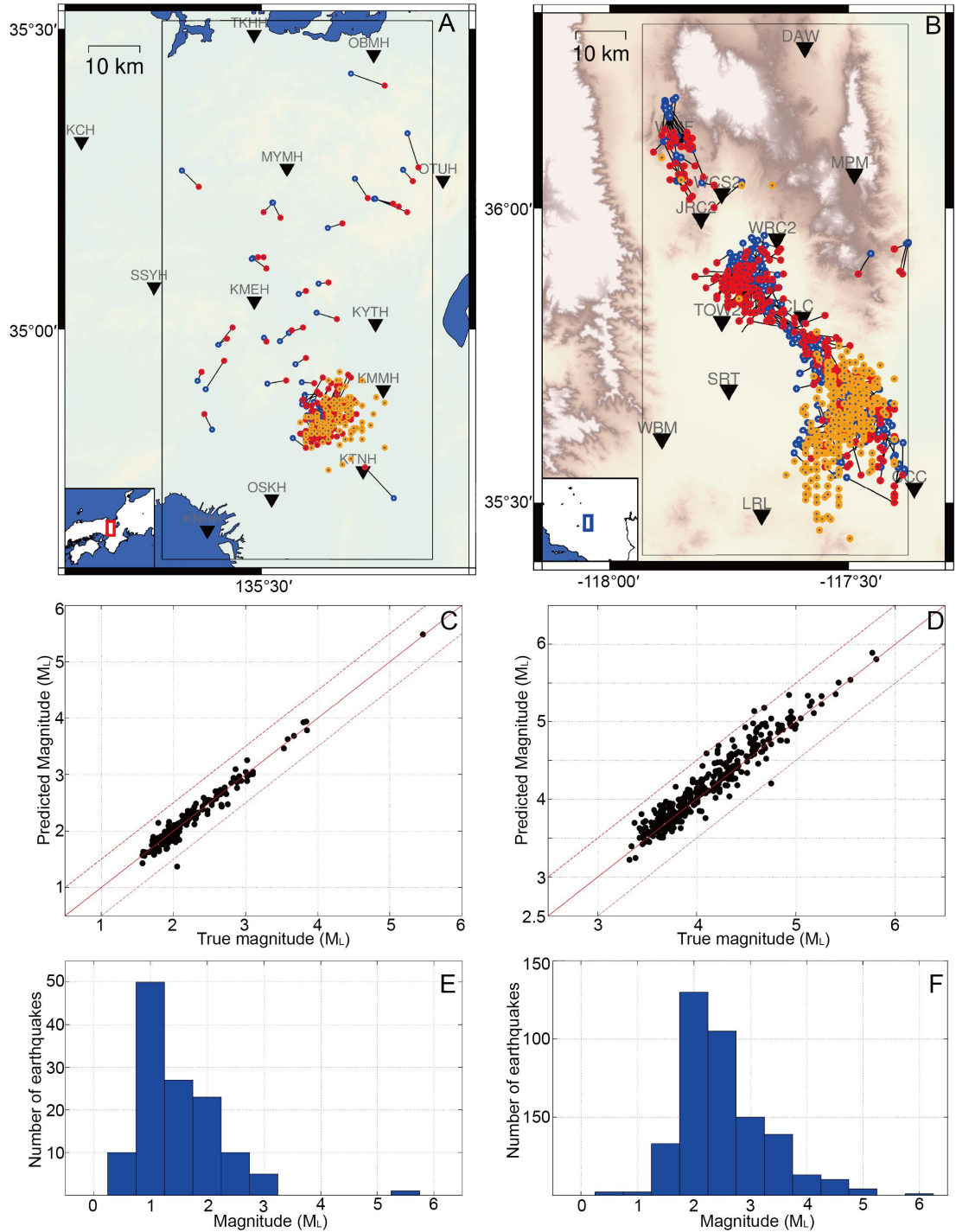

**Figure 5.** The monitoring results in Osaka, Japan (A, C, E) and Ridgecrest, US (B, D, F). (A) The predicted results of 179 cataloged earthquakes between June 16 and December 25, 2018, and 126 detected earthquakes (orange) from one-day continuous data on June 18, 2018, in Osaka,

Japan. The predicted (red) and cataloged (blue) earthquake locations are connected for comparison. (B) The predicted results of 349 cataloged earthquakes between July 4, 2019 and November 16, 2020, and 389 detected earthquakes (orange) from one-day continuous data on July 4, 2019, in Ridgecrest, US. The predicted (red) and cataloged (blue) earthquake locations are connected for comparison. (C) The magnitude comparison between the cataloged and predicted results of the 179 earthquakes in Osaka, Japan. (D) The magnitude comparison between the cataloged and predicted results of the 349 earthquakes in Ridgecrest, US. (E) The magnitude statistics of the one-day monitoring in Osaka, Japan. (F) The magnitude statistics of the one-day monitoring in Ridgecrest, US.

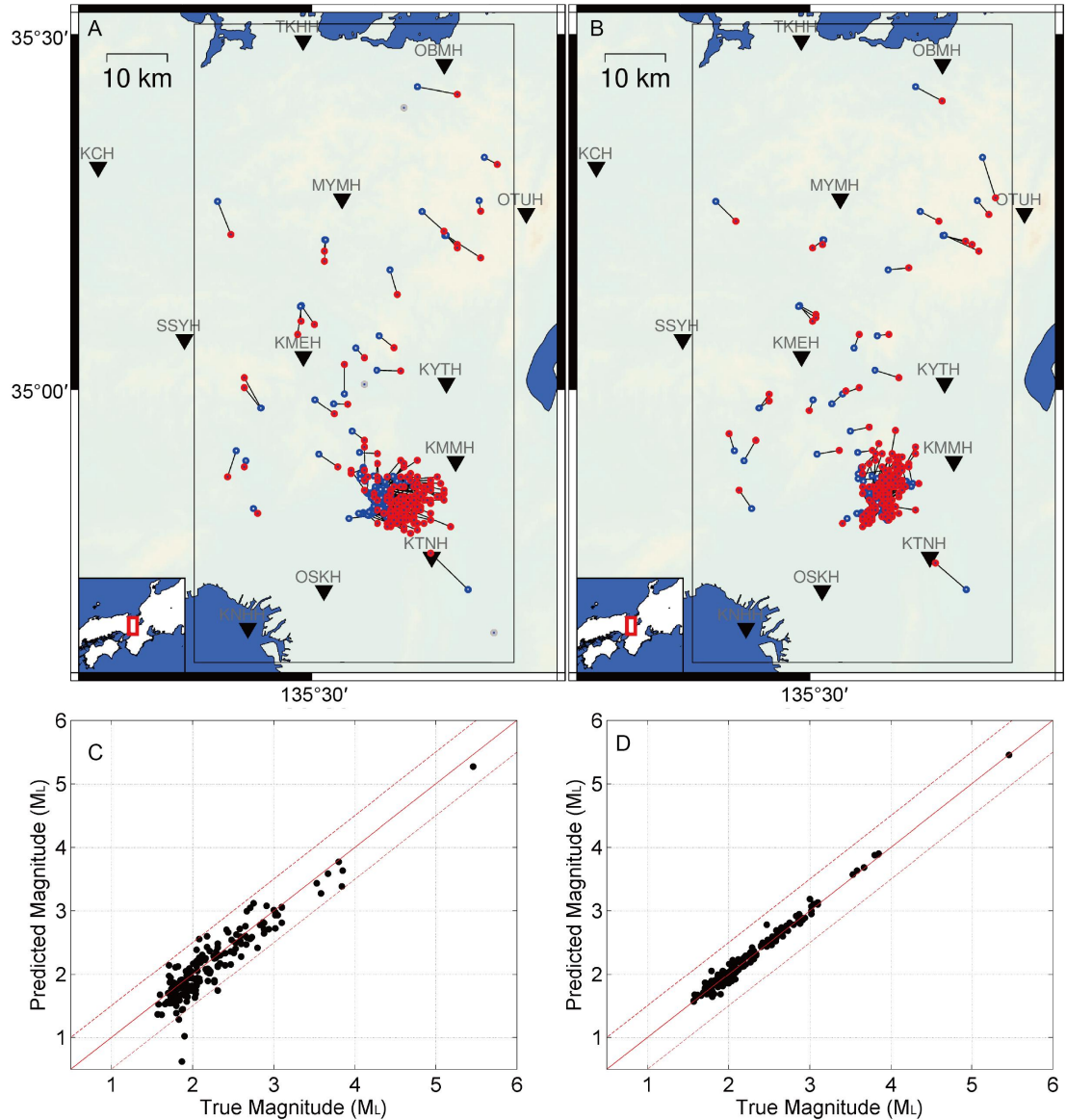

**Figure 6.** The earthquake parameter comparisons between the predicted results and the cataloged results of 179 earthquakes in Osaka, Japan. (A) The earthquake location comparison at the 4th second after the first triggered station. (B) The earthquake location comparison at the 15th second after the first triggered station. The predicted locations (red) are connected with the cataloged locations (blue), while the gray points represent predicted locations with errors larger than 15 km due to low PDFs for current time windows. (C) The earthquake magnitude

comparison at the 4th second after the first triggered station. (D) The earthquake magnitude comparison at the 15th second after the first triggered station.

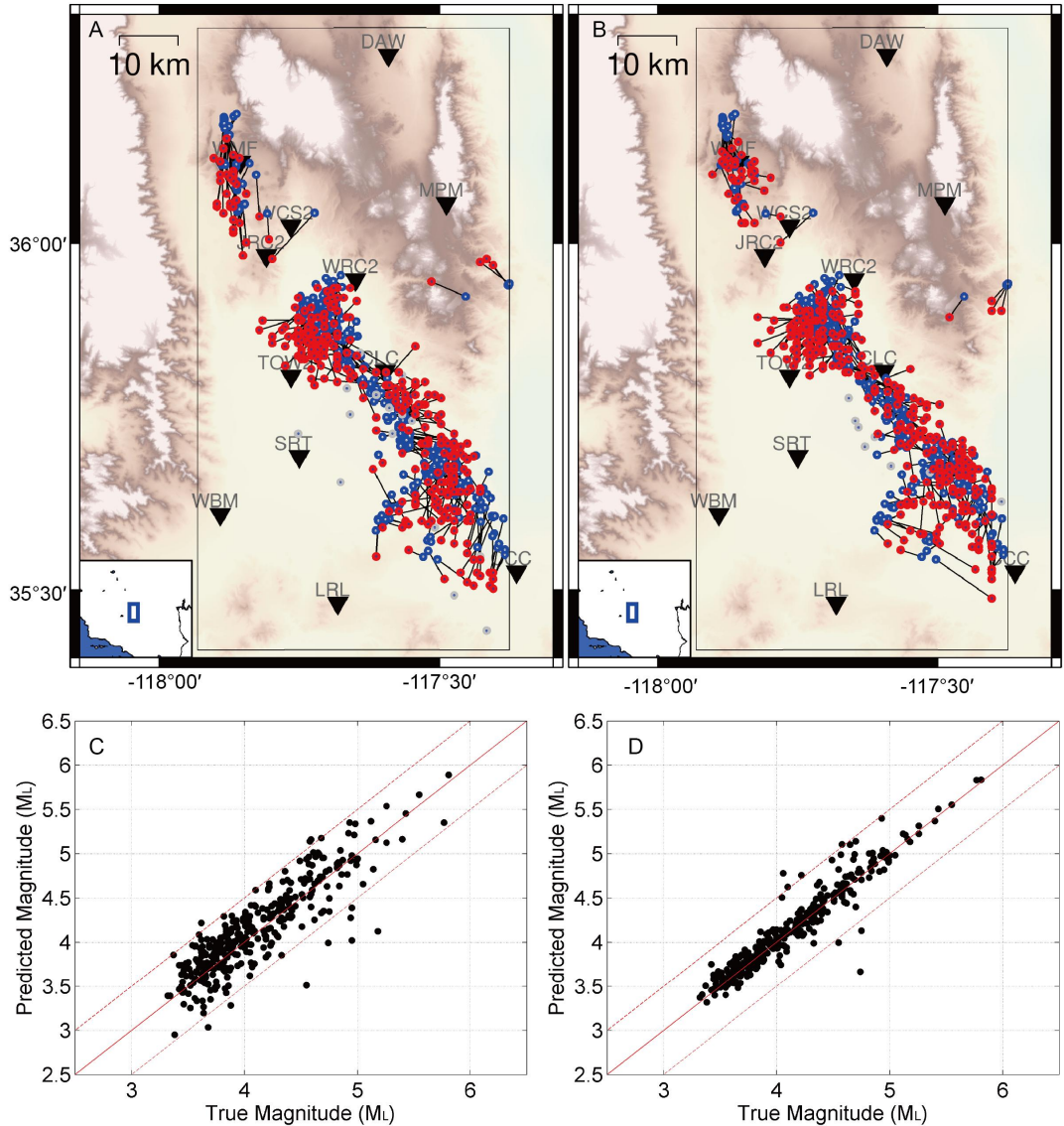

**Figure 7.** The earthquake parameter comparisons between the predicted results and the cataloged results of 349 earthquakes in Ridgecrest, US. (A) The earthquake location comparison at the 4th second after the first triggered station. (B) The earthquake location comparison at the 15th second after the first triggered station. The predicted locations (red) are connected with the cataloged locations (blue), while the gray points represent predicted locations with errors larger

than 15 km due to low PDFs for current time windows. (C) The earthquake magnitude comparison at the 4th second after the first triggered station. (D) The earthquake magnitude comparison at the 15th second after the first triggered station.

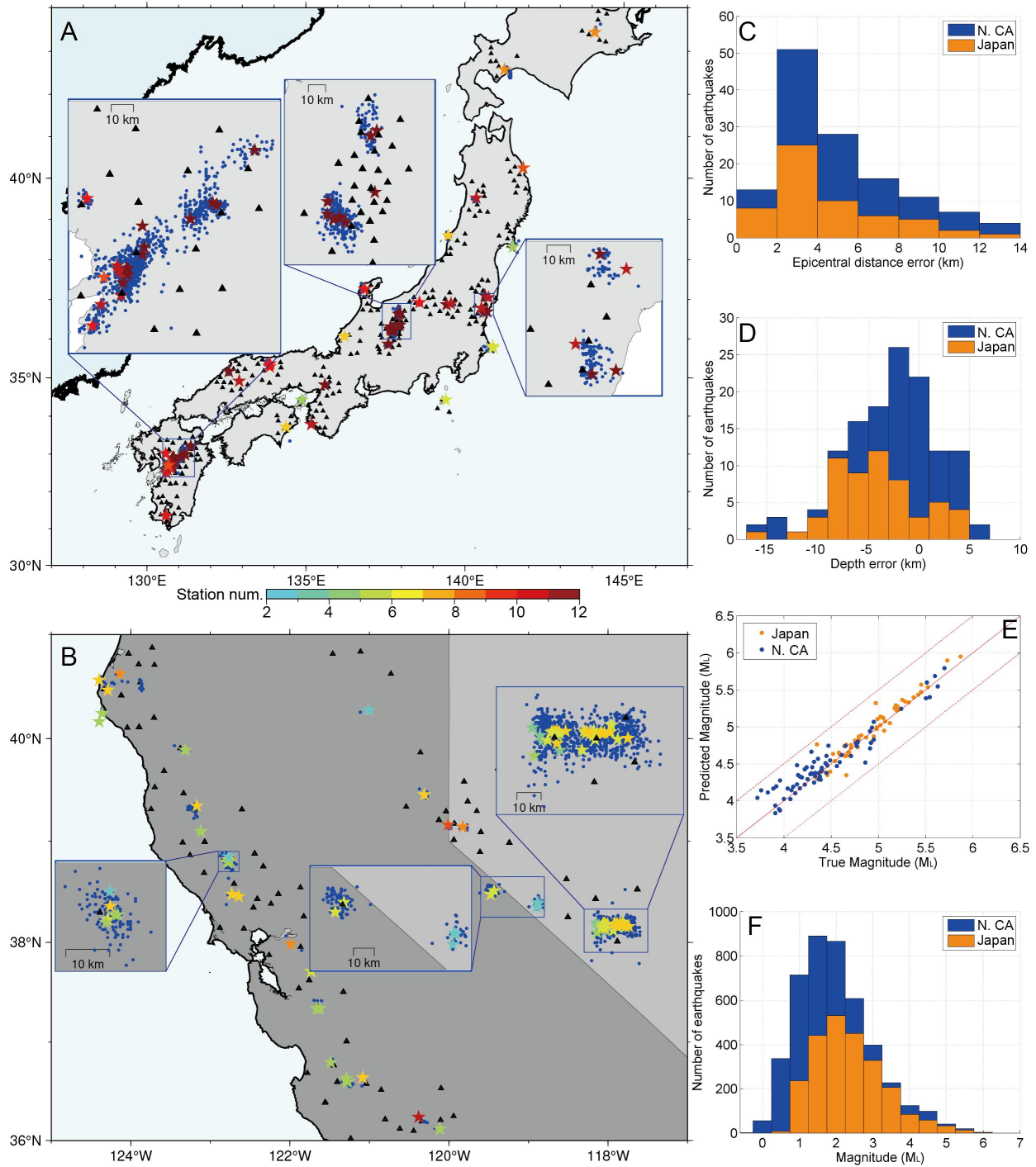

**Figure 8.** The real-time monitoring of the neural networks to the 130 relatively large earthquakes occurred in different regions. (A) Fifty-seven $M_{JMA} \geq 5.0$ earthquakes in Japan. (B) Seventy-three $M \geq 4.0$ earthquakes in North California, US. The stars are the predicted locations with the color

indicating the number of monitoring stations, and the blue dots are the detected earthquakes from one-hour continuous data following the large earthquakes. The black triangles are the selected monitoring stations for the earthquakes. (C) Epicentral distance errors. (D) Depth errors. (E) The comparison between the predicted and true magnitudes. (F) The magnitude statistics of the detected earthquakes for both regions.

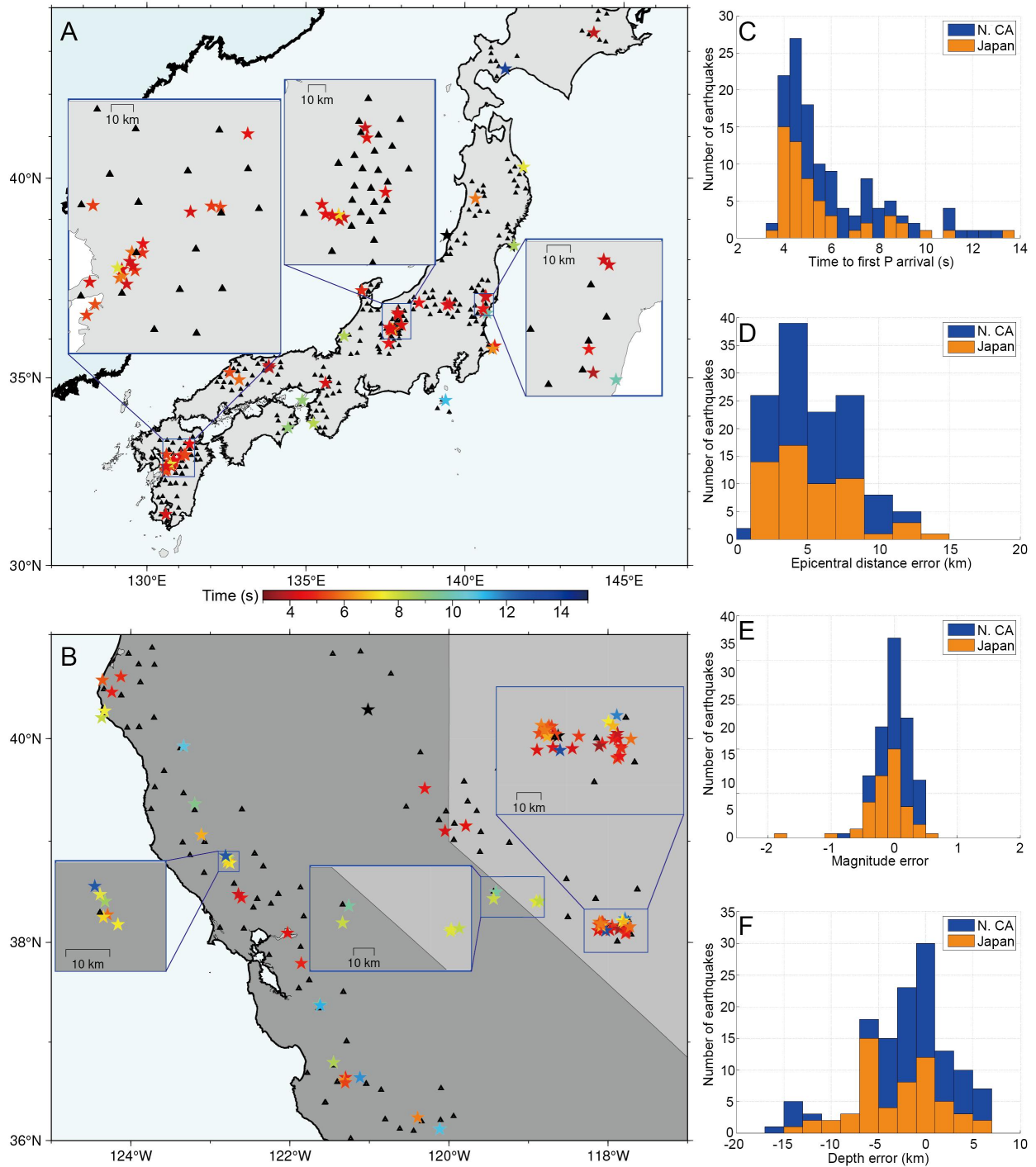

**Figure 9.** Predicted earthquake parameters at the first alarms for the 130 relatively large earthquakes. (A) Time distribution of the first alarms in Japan. (B) Time distribution of the first alarms in N. California, US. (C) The time of first alarms. The time of the first alarm is defined as the difference between the P arrival time of the first triggered station and the current time. (D) The epicentral error. (E) The magnitude error. (F) The depth error.

**Data and methods**

**Training and testing data background**

The training data primarily consist of central Italy earthquakes from January 15, 2010, to September 25, 2018, and Oklahoma induced earthquakes from April 16, 2013, to March 31, 2016. We collect waveform datasets from 4,285 central Italian earthquakes (M≥2.5) recorded by 12 permanent stations and from 1,265 Oklahoma earthquakes (M ≥ 3.0) recorded by 30 permanent stations (Fig. 3). Most of the earthquakes in Italy occur at depths ranging from 5-17 km, while the induced earthquakes in Oklahoma occur at comparably shallow depths, ranging from 0-8 km. The total number of seismograms for the Italy dataset ideally should be 51,420 (12 × 4,285), while the Oklahoma dataset should have 37,950 (30 × 1,265) seismograms. However, due to missing data in some stations and events, the final number of seismograms is 42,945 for Italy and 36,203 for Oklahoma. The corresponding epicentral distances of the collected seismograms are mostly in the range of 0-110 km. To further increase the diversity of the seismograms, we collect data from 768 earthquakes (M≥2.5) that occurred before July 4, 2019, in Southern California (Fig. 3). However, many of the datasets from the stations are missing, and we are only able to collect 15,438 training seismograms. Therefore, the total number of collected seismograms is 94,586 (42,945+36,203+15,438), and the depth and epicentral distance distributions are shown in Fig. 2B. All waveforms are resampled to 20 Hz, and the three components are rotated to the E, N, and Z directions. Additionally, the instrument responses are removed.

To simulate the early warning process, we truncate the waveforms into different time windows to mimic the scenario when there are only partial earthquake signals from the triggered monitoring stations. We randomly cut the waveform of an earthquake sample starting from 1.0 to

26 s relative to the first triggered P phases among all stations, with a total truncating window of 30 s. We randomly generate 355,001 earthquake samples across the monitoring area with a varying number of monitoring stations distributed within a specific range. We also produce an additional 2,000 earthquake samples outside of the monitoring area, which are intentionally labeled with a probability density function (PDF) of zero to represent abnormal events. These samples are then utilized for training the detection and location networks.

To evaluate the real-time monitoring efficiency and generalization ability of the trained network models, we apply them to the 2018 Osaka $M_{JMA}$ 6.1 earthquake sequence in Japan, the 2019 Ridgecrest earthquake sequence in the US, and 130 relatively large earthquakes across Japan and Nornthern California. On June 18, 2018, at 07:58 Japan Standard Time, a magnitude 6.1 earthquake struck the Osaka metropolitan area, which has a population exceeding eight million. The earthquake generated strong ground motions and resulted in four casualties. The $M_w$ 6.4 earthquake sequence that occurred on 4 July 2019 also caused damages to at least 100 homes and businesses in the communities of Ridgecrest and Trona. For the Osaka earthquake sequence on June 18, 2018, we download one day's continuous waveform recorded by 12 stations within a range of 80 km to 100 km to test the efficiency of the neural networks. In addition, we select 179 earthquakes ($M_{JMA}$ >1.0) that occurred between June 16 and December 25, 2018, to assess the accuracy of the predicted earthquake parameters by generalized neural networks. According to the routine catalog provided by the Japan Meteorological Agency (JMA), the aftershocks following the main shock of $M_{JMA}$ 6.1 are relatively weak, with magnitudes not exceeding $M_{JMA}$ 4.1. We primarily focus on the early warning of relatively large earthquakes, although the neural network is also capable of detecting smaller earthquakes. For the Ridgecrest earthquake sequence, we also download continuous seismic data recorded by 12 nearby stations on July 4,

2019, and 349 relatively large earthquakes (M≥2.5) from July 4, 2019, to November 16, 2020, to assess the performance of the neural networks. In the downloaded dataset, some stations are malfunctioning or missing; nevertheless, the neural network is capable of handling datasets with a flexible number of stations (≤12).

To assess the generalization ability, we apply the trained models to 57 relatively large earthquakes ($M_{JMA} \geq 5.0$) in Japan and 73 relatively large earthquakes (M ≥ 4.0) in North California, US. We download one-hour continuous data for each earthquake in both regions, recorded by a varying number of stations. The waveforms of the stations with epicentral distances less than 110 km are downloaded. We select the stations without excessive missing data and components, and set the monitoring area for each earthquake separately to ensure that the station range includes as many stations as possible. If the number of stations exceeds 12, we select the closest 12 stations to the epicenter for testing.

**Model architectures and training**

We design two neural networks for the detection and location of earthquakes using 2D convolutional layers, as illustrated in Fig. 1. The input for the detection neural network consists of three components of waveform data from multiple stations. The input size ($12 \times 1024 \times 3$) is determined by the maximum number of stations, the maximum length of time samples, and the number of components. Specifically, the total length of the waveform is 30 seconds, with a time interval of 0.05 seconds, resulting in 600 time samples. Any remaining 424 time samples are padded with zeroes. In our application, the number of stations can vary (less than 12), and any additional dimensions beyond station number are also padded with zeroes. The output of the neural network is labeled with a Gaussian distribution, with the peak value corresponding to the arrival time of the first triggered P phase if an earthquake event is detected. If the input contains only noise, the output is labeled with zeroes.

For the location neural network, the input includes both the XY coordinates of the stations and their corresponding waveform data. The station X and Y coordinates are normalized to a range of 0 to 1 based on the station ranges (0-82 km and 0-100 km). These normalized coordinates are represented as two vectors, each with a length of 1024 samples, occupying two channels along with the corresponding waveform data. In contrast to the graph neural network approach[28], we utilize the move-out features of the waveforms and organize the input data by sorting stations based on their X and Y coordinates in ascending order. Consequently, the total size of the input is $12 \times 1024 \times 10$, with the first 5 channels containing the three waveform components and station locations sorted by X coordinates, while the remaining 5 channels contain the same data sorted by Y coordinates. This arrangement results in that the input image pattern is unique for an earthquake monitored by the specified stations. The output of the location neural network is labeled with a 3D Gaussian distribution, with the peak value representing the predicted location. The grid size corresponds to the monitoring range of 16-66 km for X, 0-100 km for Y, and -6-22.8 km for depth. The training earthquakes are generated with depths ranging from 0 to 20 km. However, for labeling the output, we employ a broader depth range, taking into account that the Gaussian distribution radius at the boundary of deep or shallow earthquakes may exceed the depth range if the same range is utilized for labeling.

Regarding the magnitude network, we calculate the magnitude for each station individually, with an input size of $1024 \times 4$. The first three channels consist of waveform data, and the fourth channel represents the epicentral distance. Similar to the input for the detection and location neural networks, the 30-second seismic waveforms occupy 600 time samples, and any remaining 424 time samples are padded with zeros. Additionally, the epicentral distance is normalized to a range of 0-1, corresponding to distances from 0 to 110 km. During training, we randomly truncate the waveforms starting from 1 to 25 seconds relative to the P arrival time, ensuring that the final model can predict the magnitude with only a few earthquake signals within the input time window. The output of the magnitude model is labeled with a 1D Gaussian distribution, where the peak value represents a normalized magnitude. This normalization is defined by subtracting the logarithm of the maximum amplitude from the local $M_L$ magnitude, excluding the contribution of waveform amplitude[35]. This normalized magnitude allows us to standardize the neural network input with the maximum amplitude. The relationship between waveform amplitude and magnitude is explicit, with a tenfold increase in amplitude corresponding to a one-unit increase in magnitude. The final prediction of the neural network must add the logarithm of the waveform amplitude to yield the final $M_L$ magnitude.

The neural network architectures predominantly utilize Convolutional, MaxPooling, and Upsampling layers, as illustrated in Fig. 1. The design of the detection and location neural networks incorporates 2D layers to handle input data from multiple stations. In contrast, the magnitude network consists of 1D convolutional layers to process data from individual stations. The kernel sizes for the 2D and 1D convolutional layers are uniformly set to $3\times 3$ and 3, respectively. Zero-padding is applied to the output of each convolutional layer to maintain the output size.

MaxPooling layers are employed to extract critical features for constraining earthquake parameters, while Upsampling layers are used to adjust the final output size of the network model. The task of earthquake detection is relatively straightforward compared to the location and magnitude estimation. Thus, a fully convolutional neural network is sufficient to achieve the goal of earthquake detection. To mitigate the issue of gradient vanishing, we introduce multiple copy layers into the location and magnitude neural networks, as shown in Fig. 1.

The three models are trained using the Adam algorithm with a learning rate of $10^{-4}$. Additionally, they are equipped with 2, 4, and 2 Dropout layers in the detection, location, and magnitude networks, respectively[45]. The trained detection and location network models are merged into a single file, enabling simultaneous detection and location. When the merged neural network detects an earthquake event, it calculates the theoretical P arrival times. The triggered stations are determined based on whether the P arrival times fall within the monitoring time window. The waveform data from the triggered stations are then passed to the magnitude network to estimate magnitudes, and the final result is the mean value across all triggered stations.

**Acknowledgments and Data**

This work was supported by National Natural Science Foundation of China Grant (No. U2239204 to XZ), Natural Science Foundation of Jiangxi Province Grant (20224BAB211024 to XZ), and the Natural Sciences and Engineering Research Council of Canada Discovery Grant (RGPIN-2019-04297 to MZ). Seismic data in Italy and USA were downloaded through the International Federation of Digital Seismograph Networks (FDSN) web services from institutions such as Italy's National Institute of Geophysics and Volcanology (INGV), Northern California Earthquake Data Center (NCEDC), Southern California Earthquake Data Center (SCEDC). Seismic data in Japan were downloaded from NIED Hi-net operated by National Research Institute for Earth Science and Disaster Resilience. The earthquake catalog used in this study can be downloaded from http://cnt.rm.ingv.it/ (last accessed May 2023),

http://earthquake.usgs.gov/earthquakes/search/ (last accessed May 2023), https://www.data.jma.go.jp/svd/eqev/data/bulletin/index_e.html (last accessed May 2023).